\documentclass[pre,twocolumn,showpacs,superscriptaddress]{revtex4-1}
\usepackage{graphicx}
\usepackage{amsmath,amssymb}

\newcommand{\A}{\mathrm{A}}
\newcommand{\trelax}{\tau}
\newcommand{\tsamp}{t_\text{samp}}
\newcommand{\atot}{a}
\newcommand{\zmax}{z_m}

\newcommand{\myvec}[1]{{\mathbf #1}}
\newcommand{\xvec}{\myvec{x}}

\newcommand{\myav}[1]{\langle{#1}\rangle}
\newcommand{\mybigav}[1]{\Bigl\langle{#1}\Bigr\rangle}

\newcommand{\statep}{P}
\newcommand{\trajp}{{\mathbb P}}

\newcommand{\targp}{p_\text{targ}}
\newcommand{\refp}{p_\text{ref}}

\newcommand{\avtarg}[1]{\myav{#1}^{\phantom{n}}_\text{targ}}
\newcommand{\avref}[1]{\myav{#1}^{\phantom{n}}_\text{ref}}
\newcommand{\avrefbig}[1]{\mybigav{#1}_\text{ref}}

\newcommand{\targstatep}{\statep_\text{targ}}
\newcommand{\refstatep}{\statep_\text{ref}}

\newcommand{\targtrajp}{\trajp_\text{targ}}
\newcommand{\reftrajp}{\trajp_\text{ref}}

\newcommand{\targmu}{\mu_\text{targ}}
\newcommand{\refmu}{\mu_\text{ref}}

\newcommand{\latin}[1]{{\itshape #1}}
\newcommand{\etc}{\latin{et\,c.}}
\newcommand{\eg}{\latin{e.\,g.}}
\newcommand{\ie}{\latin{i.\,e.}}
\newcommand{\apriori}{\latin{a priori}}

\newcommand{\naive}{{na\"\i{}ve}}

\newcommand{\Eqref}[1]{Eq.~\eqref{#1}}
\newcommand{\Eqsref}[1]{Eqs.~\eqref{#1}}
\newcommand{\Refcite}[1]{Ref.~\onlinecite{#1}}
\newcommand{\Figref}[1]{Fig.~\ref{#1}}
\newcommand{\Figsref}[1]{Figs.~\ref{#1}}

\begin{document}

\title{Trajectory reweighting for non-equilibrium steady states}

\author{Patrick B. Warren}

\email{patrick.warren@unilever.com}

\affiliation{Unilever R\&D Port Sunlight, Quarry Road East, Bebington,
  Wirral, CH63 3JW, UK.}

\author{Rosalind J. Allen}

\email{rosalind.allen@ed.ac.uk}

\affiliation{SUPA, School of Physics and Astronomy, The University of
  Edinburgh, Peter Guthrie Tait Road, Edinburgh EH9 3FD, UK}

\date{\today}

\begin{abstract}
Modern methods for sampling rugged landscapes in state space mainly
rely on knowledge of the relative probabilities of microstates, which
is given by the Boltzmann factor for equilibrium systems.  In
principle, trajectory reweighting provides an elegant way to extend
these algorithms to non-equilibrium systems, by numerically
calculating the relative weights that can be directly substituted for
the Boltzmann factor.  We show that trajectory reweighting has many
commonalities with Rosenbluth sampling for chain macromolecules,
including practical problems which stem from the fact that both are
iterated importance sampling schemes: for long trajectories the
distribution of trajectory weights becomes very broad and trajectories
carrying high weights are infrequently sampled, yet long trajectories
are unavoidable in rugged landscapes.  For probing the probability
landscapes of genetic switches and similar systems, these issues
preclude the straightforward use of trajectory reweighting.  The
analogy to Rosenbluth sampling suggests though that path ensemble
methods such as PERM (pruned-enriched Rosenbluth method) could provide
a way forward.
\end{abstract}


\maketitle

\section{Introduction}
The Boltzmann factor, which describes exactly the relative probability
of microstates at equilibrium in systems whose dynamics obeys detailed
balance, forms the cornerstone of a plethora of simulation methods in
the physical sciences. For example, the seminal Metropolis-Hastings
algorithm for Monte-Carlo simulation exploits the Boltzmann factor to
generate a trajectory of configurations which sample the
Gibbs-Boltzmann distribution \cite{FS02}.  Knowledge of the Boltzmann
factor also makes possible a host of biased sampling methods, which
allow efficient characterisation of rugged free energy landscapes
comprising multiple free energy minima separated by barriers.  In
these methods, information on a \emph{target} system of interest is
obtained by simulating a \emph{reference} system, whose microstate
probabilities are biased to be different from the target system.  The
results are corrected for the bias by reweighting with, for example, a
Boltzmann factor.  The reference system is typically easier to sample
than the target system.  Thus in umbrella sampling \cite{TV77}, an
external potential is used to coerce the reference system (or a
sequence of such systems) to sample a free energy barrier. The basis
of biased sampling schemes is the generic relation
\begin{equation}
  \avtarg{\Theta(\xvec)}=\frac{\avref{\Theta(\xvec)\,
      W(\xvec)}}{\avref{W(\xvec)}}
\label{eq:10}
\end{equation}
where $\Theta(\xvec)$ is some quantity of interest (an order parameter
for example), the brackets $\avref{\dots}$ refer to an average over
microstates $\xvec$ for the reference system, the brackets
$\avtarg{\dots}$ refer to an average for the target system, and $W$ is
a reweighting factor
\begin{equation}
W(\xvec) \propto \frac{\targstatep^\infty(\xvec)}{\refstatep^\infty(\xvec)}\,.
\label{eq:10eq}
\end{equation}
Here the ratio $\targstatep^\infty(\xvec)/\refstatep^\infty(\xvec)$ is
the relative probability of observing the microstate $\xvec$, where
$\refstatep^\infty(\xvec)$ and $\targstatep^\infty(\xvec)$ are the
steady state (superscript `$\infty$') probability distributions for
the reference and target systems respectively.  For systems whose
dynamics obeys detailed balance this ratio is given analytically by
the Boltzmann factor (up to an overall constant of
proportionality). Thus \Eqref{eq:10} provides a way to compute
averages over the target system from a simulation of the reference
system: during the simulation, one simply tracks the quantity $W$ and
uses it to reweight the average of the quantity of interest $\Theta$.
The constant of proportionality does not need to be calculated since
it cancels in \Eqref{eq:10}.

For non-equilibrium systems, whose dynamics does not obey detailed
balance, the relative probabilities of microstates are
\apriori\ unknown. Hence one has no analytical expression for the
reweighting factor $W$, precluding the straightforward use of this
type of biased sampling scheme. Sampling non-equilibrium steady states
is important in a variety of contexts, including statistical
mechanical hopping models with driven dynamics \cite{EFG95}, sheared
soft matter systems \cite{Stratford2005} and chemical models of gene
regulatory circuits \cite{ARM98, SM04, MtW09} where failure of
detailed balance is arguably responsible for some of the most
important biological characteristics \cite{WtW05}.  This has motivated
recent interest in developing efficient sampling methods for
non-equilibrium steady states \cite{WBD07, VAM+07, DWD09a, DWD09b,
  DD10} which are not based on \Eqref{eq:10}, but instead take
alternative approaches. For example, both the non-equilibrium umbrella
sampling (NEUS) \cite{WBD07, DWD09a, DWD09b, DD10} and forward flux
sampling (FFS) \cite{VAM+07, Allen2009, BAtW12} methods involve
partitioning state space via a series of interfaces and manipulating
the statistics of trajectories between interfaces to enforce sampling
of less favourable parts of the state space. While this works for
barrier-crossing problems with a well-defined order parameter, it
involves significant overhead in terms of defining the interfaces.
  
In this work, we explore the direct use of biased sampling, via
\Eqref{eq:10}, for non-equilibrium steady states.  In this approach,
the reweighting factor $W$ in \Eqref{eq:10} is computed numerically,
on the fly, using \emph{trajectory reweighting}.  This reweighting
concept is not new: it is well-established in applied mathematics
where it is known as a Girsanov transformation and the trajectory
weight is formally a Radon-Nikodym derivative.  In applied
mathematics, trajectory reweighting is is used to calculate the
probabilities of rare events \cite{AG07,Touchette2012} and for
parameter sensitivity analysis \cite{Gla90,WR16}, and in mathematical
finance the method is widely used in the context of diffusion
equations \cite{AG07}.  In chemical and computational physics, the
notion of trajectory weights is also encountered in a host of
path-ensemble methods stemming from the seminal works of Jarzynski,
and Crooks \cite{Jar97, Cro99, Sun03}.

In principle, trajectory reweighting provides a way to generalise a
plethora of biased sampling methods to non-equilibrium systems
\cite{WA12,WA12b,WA14}.  For example it has been used in the context
of Onsager-Machlup path probabilities to reweight Brownian dynamics
trajectories \cite{OM53, Adi08}, while in kinetic Monte-Carlo schemes
trajectory weights \cite{HS07} and reweighting \cite{KM08} have been
succesfully exploited for first passage time problems \cite{GRP09,
  RGP10, DRG+11, RDG+11} and steady state parametric sensitivity
analysis \cite{PA07,WA12,WR16}.

In these existing applications one is typically interested in
reweighting trajectories with a fixed (and often relatively small)
number of steps $n$.  Here we explore, using a simple example of a
birth-death process, how trajectory reweighting can be used to compute
\emph{steady state} properties. Interestingly, this approach turns out
to be closely related to the Rosenbluth scheme for sampling the
configurations of chain macromolecules \cite{RR55, FS02, BK88, Gra97,
  Gra99}.  However, we find that it suffers from the \emph{same}
practical problem that afflicts \naive\ Rosenbluth sampling, in that
the distribution of the reweighting factor can become very broad so
that microstates that are important in the target system are hard to
sample.  Moreover, we show that this problem is controlled by the
dynamics of the target system, so that it cannot be avoided by a
judicious choice of reference system.  Thus, we conclude that a
straightforward application of trajectory reweighting is unlikely to
work, except for trivial examples.  However, the analogy to Rosenbluth
sampling suggests a possible solution in a prune-and-enrich strategy,
which we suggest as a direction for future work.

In the next section we present in more detail the theory of trajectory
reweighting, making concrete the analogy to Rosenbluth sampling, and
we also explain its practical limitations.  We then illustrate these
issues using the simple case of a birth-death process, before
suggesting ways in which the limitations of the method might be
overcome.

\section{Trajectory reweighting for biased sampling of non-equilibrium 
steady states}
\subsection{Theory of trajectory reweighting}
The trajectory weight $\trajp^n(\{\xvec_i\})$ describes the
probability of observing, in a stochastic simulation, a given sequence
of $n$ microstates $\{\xvec_i\}$ where $i=0\dots n$, given that we
start in a prescribed microstate at $i=0$. This is a well defined
mathematical object \cite{AG07} which can be expressed as
\begin{equation}
\trajp^n(\{\xvec_i\}) = {\textstyle\prod_{i=1}^n}\,
p(\xvec_{i-1}\to\xvec_{i})\,,
\label{eq:4}
\end{equation}
where the $p(\xvec_{i-1}\to\xvec_{i})$ are the probabilities of the
individual transitions in the underlying simulation algorithm.  We
assume that these $p(\xvec_{i-1}\to\xvec_{i})$ are well-defined (and
known) quantities, as is the case for simulation schemes such as
kinetic Monte-Carlo, Brownian dynamics, \etc.

Knowledge of the trajectory weight, \Eqref{eq:4}, makes possible the
above-mentioned trajectory reweighting as a biasing approach applied
to dynamical trajectories
\cite{AG07,PA07,Gla90,Touchette2012,WR16,OM53, Adi08,HS07,KM08,GRP09,
  RGP10, DRG+11, RDG+11,WA12}. In detail, one simulates a reference
system that has altered transition probabilities compared to the
target system of interest. For example in a kinetic Monte Carlo
simulation of a gene regulatory network the reference system might
have different chemical rate constants to the target system, or in a
simulation of a sheared soft matter system it might have a different
shear rate. The relative weight of a given trajectory in the target
system, compared to the reference system, is given by
\begin{equation}
\frac{\targtrajp^n(\{\xvec_i\})}{\reftrajp^n(\{\xvec_i\})}
={\prod_{i=1}^n}\>
\frac{\targp(\xvec_{i-1}\to\xvec_{i})}{\refp(\xvec_{i-1}\to\xvec_{i})}\,.
\label{eq:rw}
\end{equation}
The basic idea is that one can compute averages over trajectories in
the target system, from simulations of trajectories in the reference
system, using as the reweighting factor the relative trajectory weight
given by \Eqref{eq:rw}.  

\subsection{Trajectory reweighting for sampling steady-states}
To apply this to steady-states, we are not so much interested in the
properties of short dynamical trajectories, but rather in computing
the steady-state properties for systems with rugged landscapes where
long trajectories are required for accurate sampling. To apply
standard biased sampling schemes to compute steady-state averages
using \Eqref{eq:10}, we require the reweighting factor $W(\xvec)$ in
\Eqref{eq:10eq}, which in turn requires knowledge of the steady-state
relative probabilities of microstates
$\targstatep^\infty/\refstatep^\infty$. For systems whose dynamics do
not obey detailed balance, this quantity is not known \apriori\, but
one can show that it is given by the long-trajectory limit of the
relative trajectory weight:
\begin{equation}
W(\xvec)=\lim_{n\to\infty} \>
\avrefbig{\delta(\xvec-\xvec_n)\times\frac{\targtrajp^n}{\reftrajp^n}}\,.
\label{eq:10neq}
\end{equation}
The right hand side here is the average over all trajectories
generated in the reference system which end in microstate $\xvec$, as
ensured by the $\delta$-function (for steady state problems, the
starting microstate can be left unspecified).  This result is rather
obvious, but for completeness is derived in Appendix \ref{app:form}.
In the Appendix, we only prove \Eqref{eq:10neq} up to a constant of
proportionality but we can set this constant equal to unity, without
compromising the result since it cancels out in \Eqref{eq:10}.

Since we know the transition probabilities in \Eqref{eq:rw}, we can
compute numerically, on-the-fly, the quantity $W$ in \Eqref{eq:10neq}
during a simulation of the reference system, and use it as a `slot-in'
replacement for the factor $W(\xvec)$ in \Eqref{eq:10}, in standard
biased sampling schemes.  In practice $W$ can easily be computed
on-the-fly: when a simulation step is taken from
$\xvec_{i-1}\to\xvec_{i}$, one calculates the relative transition
probabilities $\targp/\refp$ for this step (which are known for a
given simulation algorithm), and multiplies the current estimator for
$W$ by this quantity. Appendix \ref{app:alg} contains a practical
scheme for achieving this objective, based on the notion of a circular
history array.  It is important to note that typically we would not
try to record $W$ as a function of $\xvec$ since in general the space
of microstates is very large, and each individual microstate will be
visited infrequently (the birth-death process below is something of an
exception to this).  Rather we would sample both $W$ and any
quantities we are interested in (generically denoted by $\Theta$
above) at periodic intervals, and construct the right-hand average in
\Eqref{eq:10} from these sampled values using
\begin{equation}
\avtarg{\Theta}=\frac{\avref{\Theta\,W}}{\avref{W}}
=\lim_{n\to\infty}\>\frac{\avref{\Theta(\xvec_n) \times 
{\targtrajp^n}/{\reftrajp^n}}}%
{\avref{{\targtrajp^n}/{\reftrajp^n}}}\,.
\end{equation}

\subsection{Analogy to Rosenbluth sampling}
It turns out that trajectory reweighting has many commonalities with
the classical Rosenbluth scheme for sampling the configurations of
chain macromolecules \cite{RR55, FS02, BK88, Gra97, Gra99}.  In this
scheme, one `grows' new chain configurations by addition of successive
segments, in a system that is typically crowded with surrounding
molecules. At each step in the chain growth, the position (and
possibly orientation) of the new segment is biased to avoid overlap
with the surrounding molecules (which would lead to rejection of the
chain configuration). To compensate for this bias, one associates a
reweighting factor with the chain configuration; this consists of a
product of the relative weights for each chain segment, in the biased
simulation, compared to the target system. Averages over chain
configurations are then reweighted by this factor.

Returning to \Eqref{eq:rw} for the relative trajectory weight
${\targtrajp^n}/{\targtrajp^n}$, we can see directly where the analogy
with Rosenbluth sampling arises. In \Eqref{eq:rw},
${\targtrajp^n}/{\targtrajp^n}$ consists of a product, over all steps
in the trajectory, of the relative transition probabilities
$\targp/\refp$ of the same step taken in the reference and target
systems. Thus, making an analogy between generation of a trajectory
and growth of a macromolecular configuration, the computation of
${\targtrajp^n}/{\targtrajp^n}$ is equivalent to the computation of
the reweighting factor for a configuration of a chain macromolecule of
length $n$ segments. Both these cases can be considered to be examples
of {\em iterated importance sampling} schemes, in which the required
reweighting factor consists of a product of individual weighting
factors associated with a series of steps in a chain.

\subsection{Sampling inefficiency for long trajectories}
The analogy with Rosenbluth sampling suggests trajectory reweighting
is likely to suffer from a common practical problem associated with
iterated importance sampling schemes.  For example it is well-known
that Rosenbluth sampling can become inefficient when the number of
segments in the chain macromolecule becomes large \cite{BK88, Gra97,
  Gra99}.  Returning to \Eqsref{eq:rw} and \eqref{eq:10neq}, we can
view the relative trajectory weight $W$, for a given
stochastically-generated trajectory, as the product of $n$ `random'
numbers $\targp/p$. Hence the logarithm of the trajectory weight can
be viewed as the sum of a series of random numbers:
\begin{equation}
\ln W=\lim_{n\to\infty} {\textstyle\sum_{i=1}^{n}}\> \ln 
\frac{\targp}{\refp}\,.\label{eq:10ln}
\end{equation}
Although successive steps will be correlated, if $n$ is large enough
we can apply the central limit theorem to the sum of random numbers in
\Eqref{eq:10ln}.  This implies that (for large $n$), $\ln W$ will
become normally distributed \cite{Fel68} with mean $m$ and variance
$v$, both of which are proportional to $n$. Thus the reweighting
factor $W$ is expected to be log-normally distributed \cite{AB57} with
mean $e^{m+v/2}$ and variance $(e^v-1)e^{2m+v}$. As a consequence of
this, the coefficient of variation (the standard deviation divided by
the mean) of $W$, sampled over many trajectories, is expected to be
$\surd({e^v-1})\sim e^{\alpha n}$, where $\alpha$ is some constant
coefficient. As the trajectory length $n$ increases, the variability
in the reweighting factor $W$ for the sampled trajectories increases
exponentially. Since the sampled $W$ values are used to compute
averages over trajectories, this drives an exponential blow-up of the
sampling error: this is the fundamental challenge for iterated
importance sampling schemes.  Another way to think about this is that,
as the distribution of $W$ values becomes very broad, the most
important trajectories are sampled very infrequently; in fact they
become exponentially rare.  This point is demonstrated in Appendix
\ref{app:evs} by considering the extreme value statistics of the
sampled $W$ values.

\subsection{The required trajectory length is set by the target system}
It turns out that one cannot avoid sampling long trajectories when
using trajectory reweighting to sampling steady states in systems with
rugged landscapes. For example, in models for genetic switches, the
system typically undergoes stochastic flips between alternative stable
states on a timescale that is far longer than that of the underlying
molecular events \cite{SM04,MtW09,Morelli2008}.  One might hope that
by choosing to simulate a reference system whose flipping rate is much
faster than that of the target system, one could sample the entire
state space more effectively. To understand why this is \emph{not} the
case, we derive an evolution equation for the reweighting factor $W$,
in Appendix \ref{app:form}. This equation (\Eqref{eq:tq1b}) shows that
the dynamics of $W$ (and hence its relaxation time) is controlled by
the transition probabilities in the target system, not the reference
system.  Thus, even if the reference system does achieve rapid
sampling of the whole state space, the quantity that we need to sample
to compute averages in the target system ($W$ and $\Theta\,W$ in
\Eqref{eq:10}) will only converge on a timescale set by the unbiased
dynamics of the target system. This implies that long trajectories,
with their associated broad distribution of trajectory weights, are
needed to compute steady-state averages in systems with rugged
landscapes.

For an intuitive explanation of this \cite{private} consider that
trajectory reweighting should faithfully reproduce the behaviour of
the target system, \emph{including transients}.  This is exploited for
instance in the first-passage time problems mentioned in the
introduction \cite{GRP09, RGP10, DRG+11, RDG+11}.  Therefore, to
achieve steady state, one has to wait until all the transients in the
target system have decayed away.  This implies that the relaxation to
steady state is governed by the target system and not the reference
system.

\section{Example: birth-death process}
We now illustrate the use of trajectory reweighting to sample
non-equilibrium steady states, and its associated issues, using a
simple example: a toy model of a birth-death process \cite{Fel68} with
birth rate $\lambda$ and death rate $\mu$,
\begin{equation}
\emptyset\overset{\lambda\>}{\to}\A\overset{\mu\>}{\to}\emptyset\,.
\label{eq:bd}
\end{equation}
This might represent a set of chemical reactions in which molecules of
type $\A$ are created stochastically in a Poisson process with rate
parameter $\lambda$ and removed from the system stochastically with
rate parameter $\mu$.  We simulate the stochastic process represented
by \Eqref{eq:bd} using the Gillespie kinetic Monte-Carlo algorithm
\cite{Gil77}.
 
For this model, the microstate of the system is fully defined by the
(discrete) copy number of $\A$, which we denote $x$; hence
$\xvec\mapsto x=0, 1, 2\dots$. The steady-state distribution of $x$ is
given analytically by a Poisson distribution,
\begin{equation}
\statep^{\infty}(x)=\frac{k^x e^{-k}}{\Gamma(x+1)}\,,
\label{eq:poiss}
\end{equation}
where $k=\lambda/\mu=\myav{x}\equiv\sum_{x=0}^\infty x
\statep^{\infty}(x)$ is the mean copy number.  The fact that
analytical results are available for this model allows rigorous
testing of the results of our stochastic simulations with trajectory
reweighting. We define a timescale for the model by setting
$\lambda=1$. We suppose that our target system of interest has death
rate $\targmu$, and we wish to compute information about the target
system by simulating a reference system with a different death rate
$\refmu$. Trajectory reweighting is implemented in our simulations as
described in Appendix \ref{app:alg}; the reweighting factor is
computed on-the-fly using a history array of length $n$.

\begin{figure}
  \begin{center}
    \includegraphics[clip=true,width=\columnwidth]{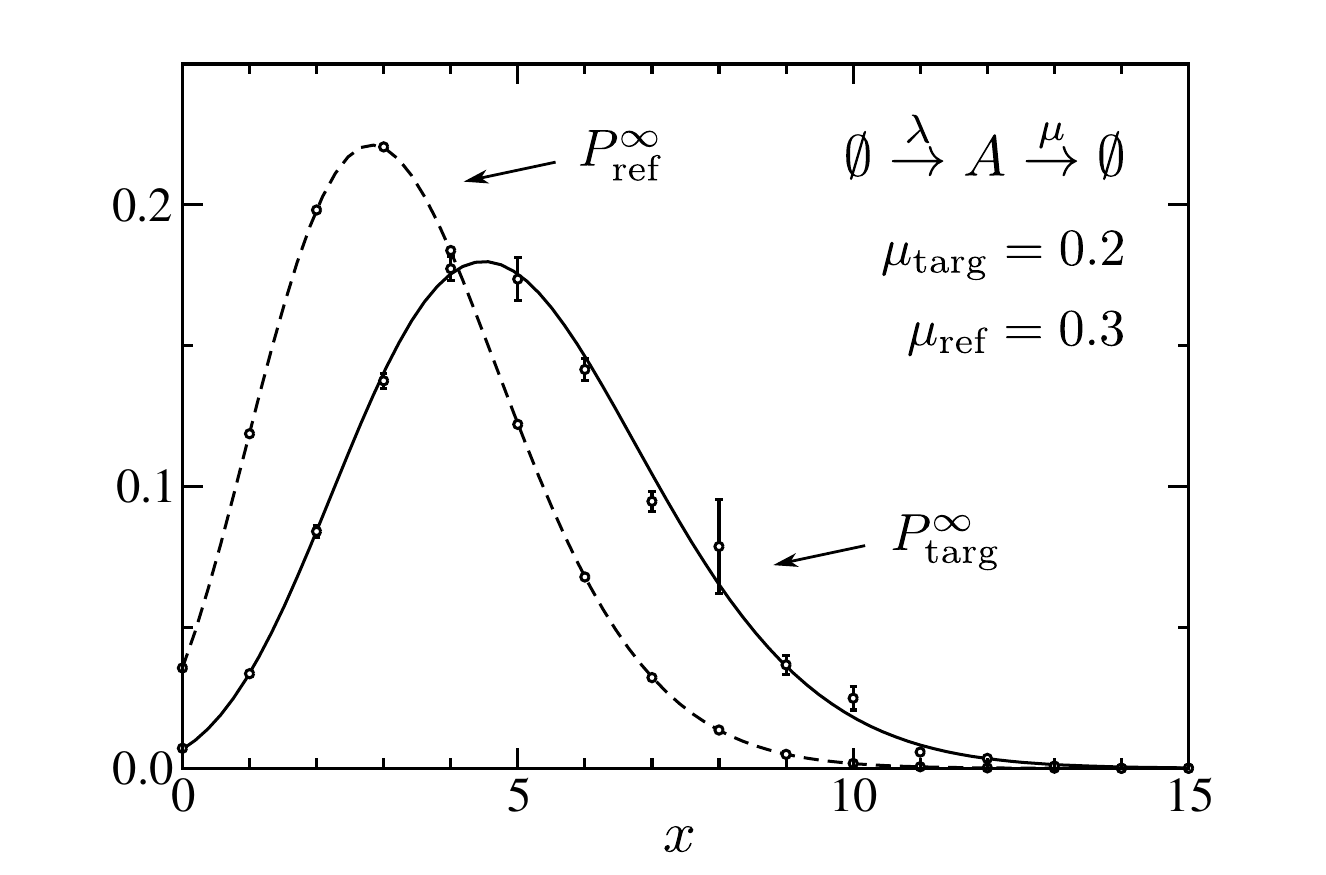}
  \end{center}
  \vskip -0.5cm
  \caption{Reweighting the birth-death process.  The steady state
    distributions for the reference and target systems are compared to
    the expected Poisson distributions from \Eqref{eq:poiss} in the
    text.  Parameters are $\lambda=1$, $\refmu=0.3$ (simulated
    reference system), $\targmu=0.2$ (target system).  The trajectory
    length was $n=50$.  Error bars (one standard deviation) are from
    block averaging (10 blocks, each of $10^5$
    samples).\label{fig:bdp}}
\end{figure}

We first demonstrate that trajectory reweighting works, in the sense
that it gives correct results for the steady-state properties of the
target system. Perhaps the most obvious system property of interest is
the mean copy number $\avtarg{x}$; this can be computed by setting
$\Theta(x)=x$ in \Eqref{eq:10}:
$\avtarg{x}=\avref{xW}/\avref{W}$. Using trajectory reweighting with
$\refmu=0.3$, $\targmu=0.2$, and $n=50$ steps, we obtain
$\avref{xW}/\avref{W}=5.03(7)$, which compares to an analytical result
of $\avtarg{x}=5$ for the target system (since
$\lambda/\targmu=5$). For the simulated reference system (for which
$\lambda/\refmu=10/3$) we obtain, as expected,
$\avref{x}=3.333(2)$. We can also examine the full probability
distribution of the copy number $\targstatep^{\infty}(x)$, which can
be obtained from \Eqref{eq:10} by setting $\Theta(x)=\delta_{xy}$
(\ie\ the Kronecker delta): $\targstatep^{\infty}(y)=
\avref{\delta_{xy}}=\avref{\delta_{xy}W}/\avref{W}$.  This should
correspond to the Poisson distribution,
\Eqref{eq:poiss}. \Figref{fig:bdp} shows our simulation results for
$\targstatep^{\infty}(x)$, compared to \Eqref{eq:poiss}, for the same
parameter set. The target system distribution obtained from our
trajectory reweighting simulation is indeed in good agreement with the
analytical result; although there is some loss of accuracy and an
increase in the sampling error in the tail of the reweighted
distribution. Plotting also the distribution $\refstatep^{\infty}(x)$
for the reference system (which also corresponds, as expected, to a
Poisson distribution, with a different parameter $k$), we see that the
increased sampling error for the target distribution occurs for
regions of state space ($x$) where the overlap between the target and
reference distributions is small; \ie\ values of $x$ which the
reference system samples poorly. This kind of problem is common to
many reference sampling schemes and is often addressed by a more
sophisticated choice of reference system; for example, as in umbrella
sampling \cite{FS02} one might use a series of reference systems
designed to split the underlying probability landscape into
subregions, each of which can be sampled more generously before being
stitched together.

\begin{figure}
  \begin{center}
    \includegraphics[clip=true,width=\columnwidth]{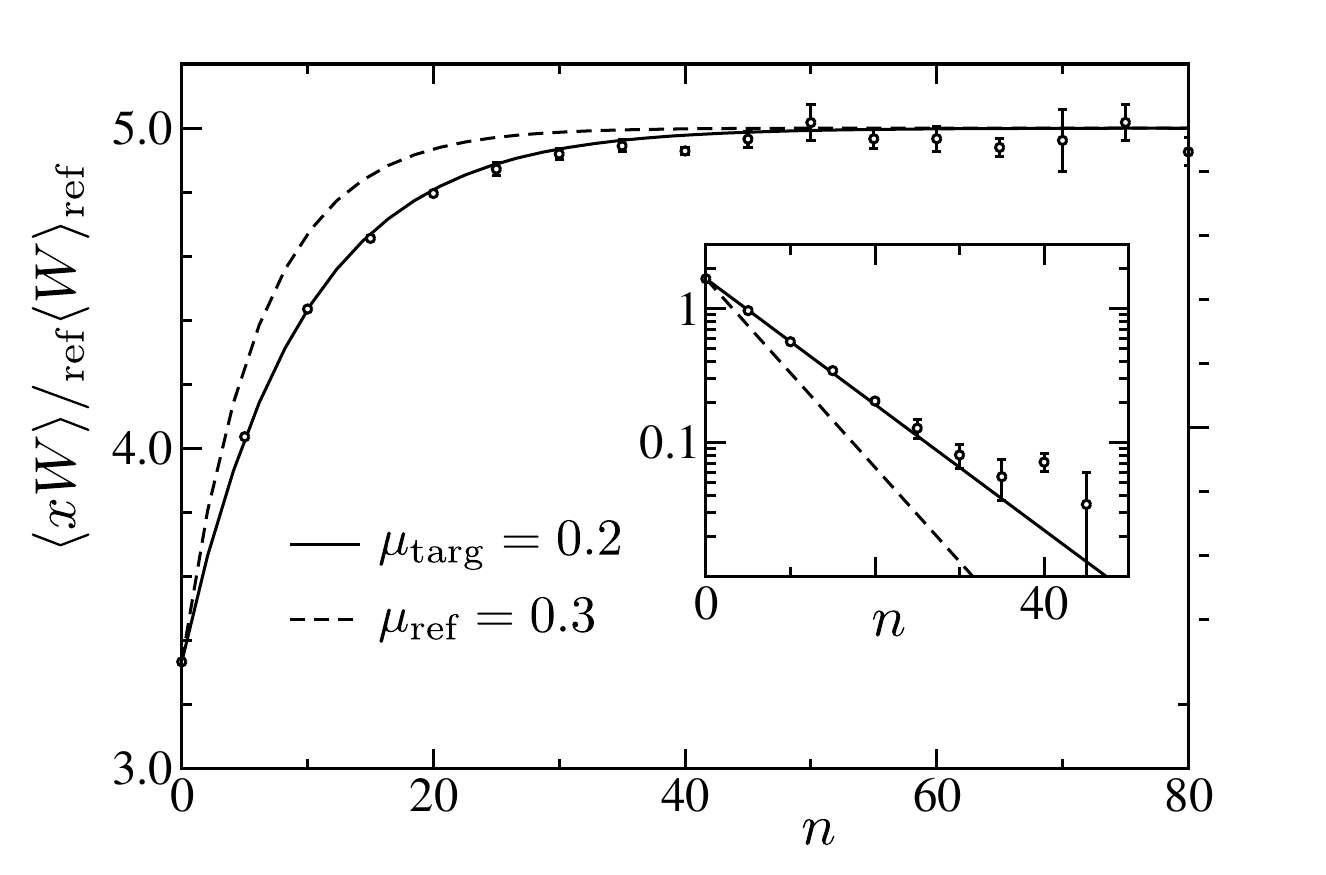}
  \end{center}
  \vskip -0.5cm
  \caption{Mean copy number in target system (points with error bars)
    as a function of trajectory length, compared to expected
    convergence rate from target (solid line) and reference (dashed
    line) systems.  The inset shows a semi-log plot of the difference
    from the expected $n\to\infty$ limit.  Each data point is a
    separate simulation along the lines of
    \Figref{fig:bdp}.\label{fig:conv}}
\end{figure}

We next illustrate the fact that the length of trajectory $n$ that is
needed for accurate sampling is governed by the target system, not the
reference system.  To this end, again using reference and target
system parameters $\refmu=0.3$ and $\targmu=0.2$, we vary the size of
the history array, \ie\ the length of trajectory $n$ that is used in
the computation of the reweighting factor $W$. \Figref{fig:conv} shows
results for the mean target system copy number, computed as
$\avtarg{x}=\avref{xW}/\avref{W}$, as a function of $n$.  When the
stored history array is very short, the method gives incorrect results
(\eg\ when $n=0$, then $W=1$ everywhere by definition, and
$\avref{xW}/\avref{W}=\avref{x}=10/3$).  As $n$ increases,
$\avref{xW}/\avref{W}$ converges to the correct result $\avtarg{x}=5$
for the target system.  We can probe the rate of this convergence by
fitting the data in \Figref{fig:conv} to a mono-exponential
relaxation,
\begin{equation}
\frac{\avref{x W}}{\avref{W}} = \frac{\lambda}{\targmu} + 
\Bigl(\frac{\lambda}{\refmu}-\frac{\lambda}{\targmu}\Bigr)
\exp\Bigl(-\frac{n\avref{\delta t}}{\trelax}\Bigr)\,,
\label{eq:rate}
\end{equation}
where $\avref{\delta t}=0.540$ is the mean simulation time step
\cite{deltat}.  \Figref{fig:conv} shows this relation plotted for both
$\trelax=\targmu^{-1}$ (solid line) and $\trelax=\refmu^{-1}$ (dotted
line), corresponding to the characteristic relaxation timescales of
the target and reference systems respectively.  The data clearly fits
$\trelax=\targmu^{-1}$, rather than $\trelax=\refmu^{-1}$, confirming
that it is the target system, not the reference system, whose dynamics
controls the convergence rate.

\begin{figure}
  \begin{center}
    \includegraphics[clip=true,width=\columnwidth]{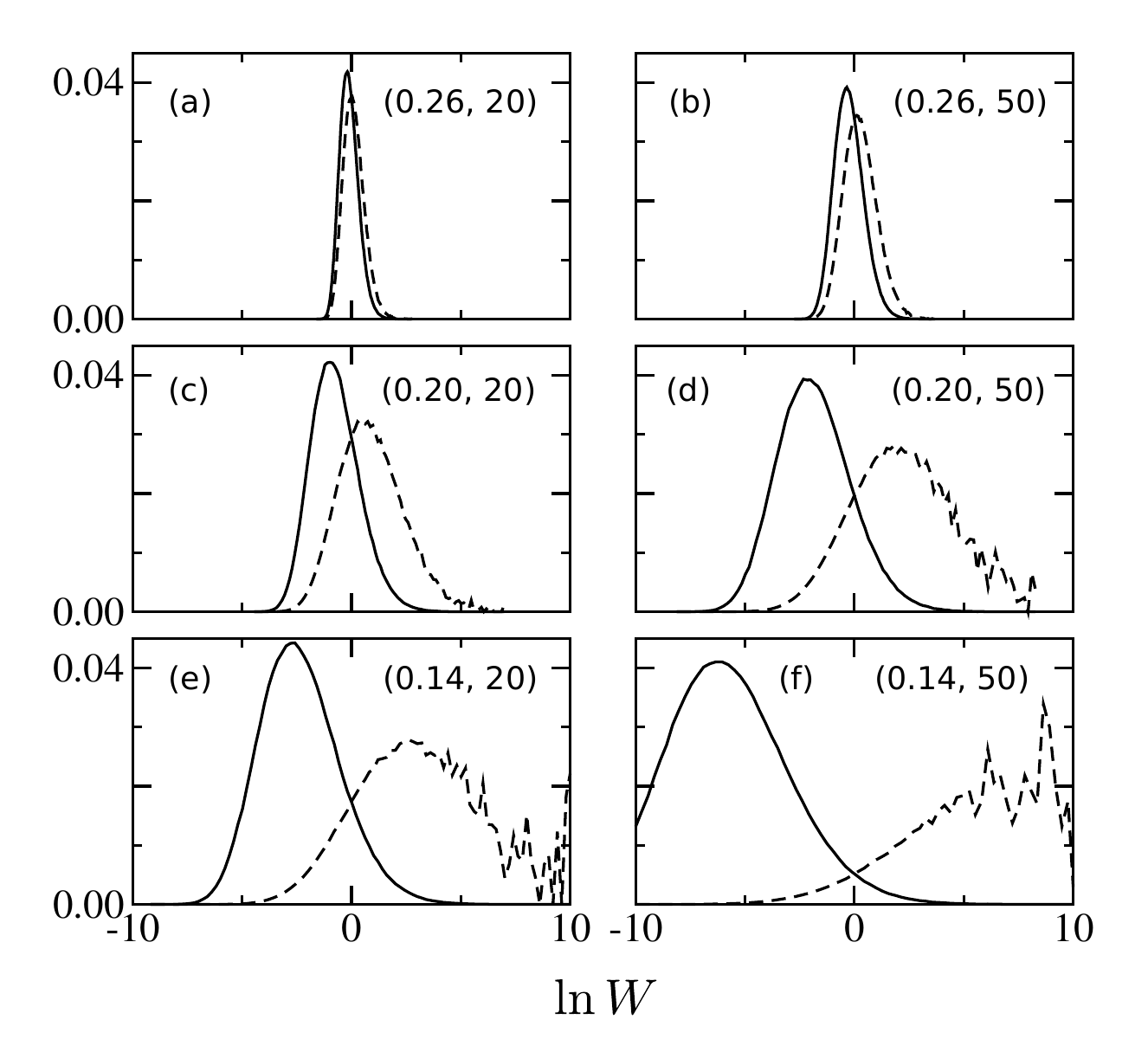}
  \end{center}
  \vskip -0.5cm
  \caption{The functions $P(\ln W)$ (solid lines) and $W\times P(\ln
    W)$ (dashed lines) for $n=20$ (a, c, e) and $n=50$ (b, d, f); and
    for $\targmu=0.26$ (a, b), $\targmu=0.20$ (c, d), and
    $\targmu=0.14$ (e, f).  Panels are also labelled directly by
    ($\targmu$, $n$), and these values of $\targmu$ are also shown
    arrowed in \Figref{fig:varymu}.\label{fig:plw}}
\end{figure}

We now test our prediction that, as the length $n$ of the trajectory
used to compute the reweighting factor $W$ increases, the distribution
of $W$ values sampled will broaden, making it hard to compute accurate
results. \Figref{fig:plw} (solid lines) shows histograms for $\ln W$
computed from our simulations for $n=20$ (left panels) and $n=50$
(right panels). As predicted by our theoretical analysis, $P(\ln W)$
indeed approaches a normal distribution for large values of $n$ and
this distribution indeed broadens as $n$ increases. Comparing results
for different values of the target system death rate $\targmu$ (top to
bottom in \Figref{fig:plw}), we see that the distribution $P(\ln W)$
also broadens as $\targmu$ decreases, \ie\ as the target and reference
system become more different from each other.  This is because the
distribution of $\targp/\refp$ values becomes wider as the reference
system deviates further from the target system.

When computing weighted averages (as in \Eqref{eq:10}) what is
important is actually $W\times P(W)$, or $W\times P(\ln W)$
\cite{wplnw} since trajectories with high weight $W$ contribute more
to the average.  We therefore also plot in \Figref{fig:plw} (dashed
lines) the distribution of values of $W\times P(\ln W)$.  Comparing
the results for different trajectory lengths $n=20$ (left panels) and
$n=50$ (right panels), we see that as the trajectory length increases,
not only does $P(\ln W)$ broaden (solid lines), but also the peak in
$W\times P(\ln W)$ gets shifted further into the tails of $P(\ln
W)$. Thus for longer trajectories we become less likely to sample the
relevant parts of $W\times P(W)$, where the weight $W$ is large. We
also see the same effect upon decreasing $\targmu$ (top to bottom
panels in \Figref{fig:plw}); as the target and reference systems
become more dissimilar, sampling trajectories with a high weight in
the target system becomes less likely \cite{poor}.  \Figref{fig:plw}
rather dramatically underscores the importance of the extreme high
weight trajectories in calculating weighted averages.  Exactly the
same kind of phenomenology is seen for Rosenbluth weights \cite{BK88}.
Indeed, \Figref{fig:plw} closely mirrors Fig.~2 in \Refcite{Gra99} for
polymers localised in random media.

Figure \ref{fig:varymu} illustrates how the difficulty in sampling the
relevant parts of $W\times P(W)$ leads to poor statistics when
computing averages in the target system. Here, we plot trajectory
reweighting simulation results for the ratio of the computed value of
$\avref{xW}/\avref{W}$, to the analytical result
$\avtarg{x}=\lambda/\targmu$ (note that since $\lambda=1$ this ratio
is equal to $\targmu \times \avref{xW}/\avref{W}$). As in
\Figref{fig:plw}, results are shown for two values of the trajectory
length, $n=20$ (dashed line) and $n=50$ (solid line), and for various
values of $\targmu$. We first note that when $\targmu$ is small (slow
relaxation of the target system), there is a systematic deviation from
the correct result $\targmu \times \avref{xW}/\avref{W}=1$; this is
because $\avref{xW}/\avref{W}$ has not yet reached steady
state. Increasing the trajectory length $n$ from 20 to 50 indeed
decreases this deviation. However, we pay a penalty for this in terms
of the sampling error; the error bars on our data points become much
larger for the longer trajectories.  This reflects the increasingly
poor sampling of the underlying $W$ distribution
(\Figref{fig:plw}). If we attempt to further reduce the systematic
error for small values of $\targmu$ by increasing $n$ still further,
for example to $n=100$, then we see so few of the important but
exponentially rare high-weight trajectories that the results become
statistically meaningless.  This illustrates rather clearly the
`Catch-22' practical issue associated with trajectory reweighting for
computing steady-state properties: for target systems with slow
dynamics, long trajectories are needed, but because trajectory
reweighting is an iterated importance sampling scheme, its statistical
accuracy decreases, catastrophically, as the trajectory length
increases.

\section{Discussion}
\begin{figure}
  \begin{center}
    \includegraphics[clip=true,width=\columnwidth]{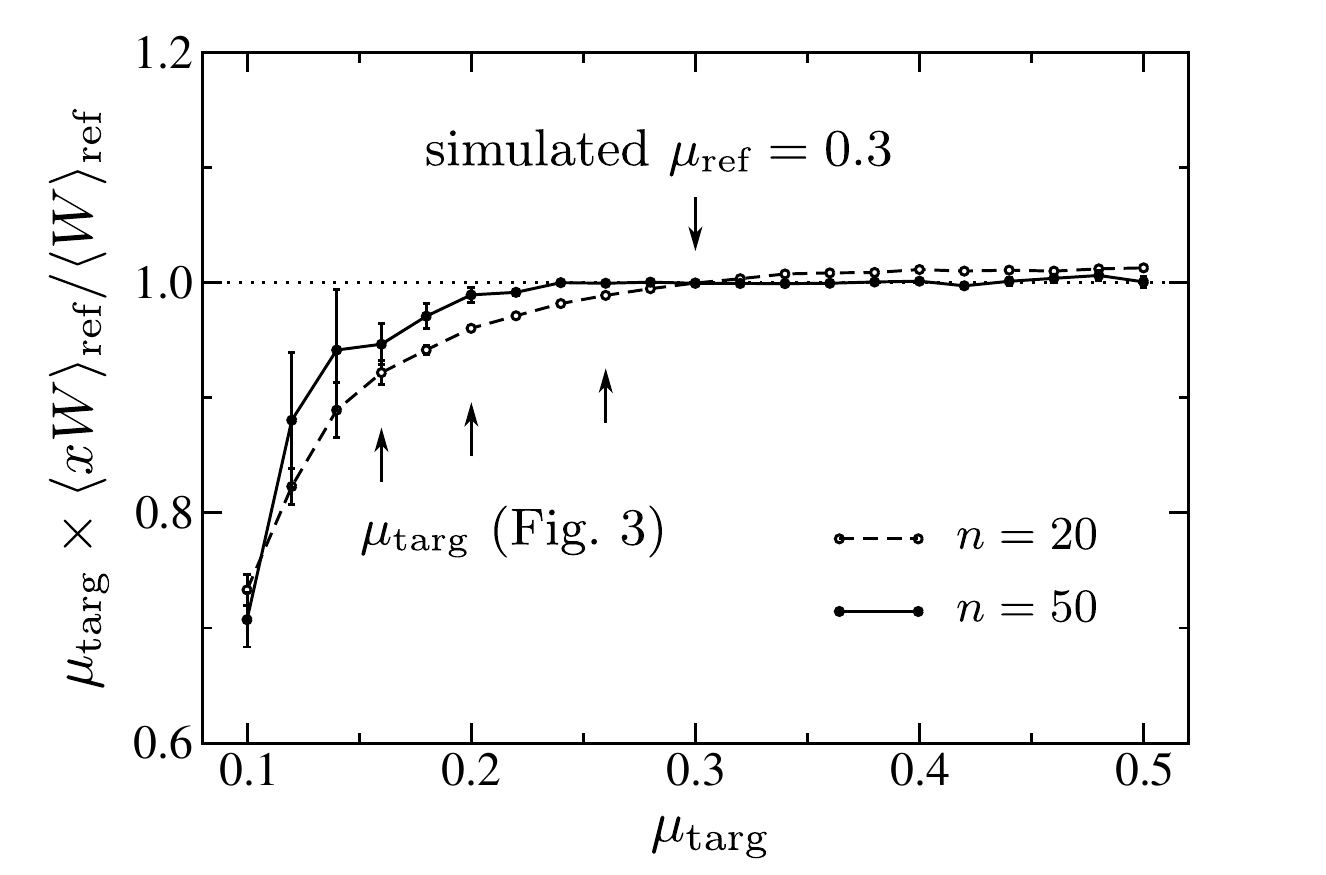}
  \end{center}
  \vskip -0.5cm
  \caption{The ratio of the weighted estimate of the target mean copy
    number to the expected value, as $\targmu$ varies, at fixed
    $\refmu=0.3$ (this quantity should $\equiv1$). Each data point is
    a separate simulation along the lines of \Figref{fig:bdp}.  Two
    values of the trajectory length are explored, and the
    upward-pointing arrows indicate the values of $\targmu$ used in
    \Figref{fig:plw}.\label{fig:varymu}}
\end{figure}

Trajectory reweighting provides an apparently elegant way to extend
biased sampling methods developed for equilibrium steady states, to
non-equilibrium systems whose dynamics do not obey detailed
balance. By simulating a reference system, which is easier to sample
than the target system of interest, one can obtain averages over the
(un-simulated) target system by reweighting averages in the simulated
reference system using reweighting factors computed from the reference
system trajectories.  The analysis presented here shows this is
possible in principle, and that it does indeed work for the
not-too-challenging test case of a birth-death process.  For this
particular problem though, it is clear that trajectory reweighting
will never beat straightforward sampling: if the target system relaxes
more slowly than the reference system ($\targmu\alt\refmu$ in
\Figref{fig:varymu}) one comes up against the problem of poor sampling
of the high-$W$ trajectories (\Figref{fig:plw}); conversely if the
target system relaxes faster than the reference system
($\targmu\agt\refmu$) it is simply more efficient to simulate the
target system directly.  However, we expect trajectory reweighting has
the potential to be useful for systems whose dynamics makes slow
switches between alternative ``basins of attraction'' (\eg\ a genetic
switch), since using a reference system that relaxes faster than the
target system should allow better sampling of trajectories that switch
between the basins.

Our work reveals the two significant practical issues which preclude
the straightforward application of trajectory reweighting.  First, the
time scale to obtain unbiased estimates of reweighted quantities is
still set by the \emph{target system}.  For example, to calculate
steady state averages in our toy model, one needs trajectory durations
which are 2--3 times the relaxation time in the target system
($\targmu^{-1}$ ; see \Figsref{fig:conv} and \ref{fig:varymu}).  We
expect this rule-of-thumb to hold generally.  Second, for long
trajectories, the distribution of trajectory weights becomes extremely
broad, and the weighted averages that we need to compute are sensitive
to the rarely visited high weight region.  In our toy model this is
exemplified in \Figref{fig:plw}. This problem becomes acute for target
systems which involve barrier-crossing events, such as genetic
switches. Since the barrier-crossing frequency is very small, the
relaxation times are very long.  Then, we cannot escape from the fact
that very lengthy trajectories are required to compute steady state
averages in such target systems, even if the reference system
equilibrates rapidly.

It is important to note that this problem is specific to the use of
trajectory reweighting for {\em steady-state} system
properties. Trajectory reweighting has been successfully applied in
numerous other contexts, including financial mathematics \cite{AG07},
rare event analysis in applied mathematics \cite{AG07,Touchette2012},
and first passage time problems \cite{GRP09, RGP10, DRG+11, RDG+11}.
The essential difference is that these problems are characterised by
\emph{short-duration} trajectories.  For sampling of short
trajectories, the measurement statistics of infrequent events can be
drastically improved by generating a large number of successful but
biased trajectories, even though these may carry a low weight once the
bias is removed.  This variance-reduction mechanism was cogently
argued by Gillespie, Roh and Petzold \cite{GRP09}.  In contrast, to
sample steady-state properties, \emph{long-duration} trajectories have
to be reweighted, with the associated sampling problems discussed
here.

We have noted that there is a close analogy between computing
trajectory weights, and Rosenbluth sampling for chain macromolecules.
This suggests a possible way forward.  In the Rosenbluth method,
inefficient sampling of high-weight polymer configurations was
resolved by the developent of the pruned-enriched Rosenbluth method
(PERM) \cite{Gra97, FS02}, and its descendant flat-histogram PERM
\cite{PK04, BAtW12}. To apply this in the present situation one would
follow an \emph{ensemble} of trajectories, and in order to keep the
weights in the desired region in $P(\ln W)$ (\Figref{fig:plw}), at
intermediate times discard trajectories with low weights, and clone
trajectories with high weights.  The added complications are the
increased bookkeeping required to keep track of a variable number of
trajectories, and the need to fine-tune the hyper-parameters in the
algorithm.  We note that PERM applied to this problem does not
necessarily avoid the issue that the sampling time is set by the
target system, thus for instance for a genetic switch we expect one
would need to sample for 2--3 times the waiting time to cross the
barrier.  However by diminishing or removing the barriers in a rugged
probability landscape, we expect that the use of trajectory
reweighting will distribute the sampling effort more effectively over
state space, leading to more accurate calculation of steady state
probability distributions.  Undoutedly this will only become clear
with further and more detailed investigations, which we leave for
future work.  Another possible avenue to explore is the use of
simultaneous trajectory reweighting across multiple systems, as in the
extended bridge sampling method \cite{MC09}.

We thank H.\ Touchette for discussions and a critical reading of an
earlier version of this manuscript, D.\ Frenkel for drawing attention
to the analogy to Rosenbluth weights, and T.\ Prellberg for
highlighting to us the advanced PERM-based methods for polymer
sampling. RJA was supported by a Royal Society University Research
Fellowship and by EPSRC under grants EP/I030298/1 and EP/J007404/1.
PBW thanks the Higgs Centre for Theoretical Physics for funding travel
under the Associate Member program.  PBW discloses a substantive ($>$
\$10k) stock holding in Unilever PLC.

\appendix

\section{Formalism for trajectory reweighting}\label{app:form}
Here, we derive \Eqref{eq:10neq} and show formally that the relaxation
time of the reweighting factor $W$ is controlled by the dynamics of
the target system rather than the reference system. We define
$\targstatep^n(\xvec)$ and $\refstatep^n(\xvec)$ as the probabilities
that, respectively, the target and reference system are in microstate
$\xvec$ after $n$ steps.  We can leave the starting state unspecified.
The steady-state probability distributions $\targstatep^\infty(\xvec)$
and $\targstatep^\infty(\xvec)$ in the main text are the large $n$
limits of $\targstatep^n(\xvec)$ and $\targstatep^n(\xvec)$.  In
addition to these we define
\begin{equation}
W^n(\xvec) = \avrefbig{\delta(\xvec-\xvec_n) \times 
\frac{\targtrajp^n}{\reftrajp^n}}\,.
\label{eq:q1}
\end{equation}
This is the mean weight per trajectory of length $n$ that ends in
$\xvec$.  In the limit $n\to\infty$, it is the quantity that features
in the right hand side of \Eqref{eq:10neq} in the main text.

The actual weight in microstate $\xvec$ is obtained by multiplying
$W^n(\xvec)$ by the probability that a trajectory ends in microstate
$\xvec$, namely $W^n(\xvec)\times\refstatep^n(\xvec)$.  An evolution
equation for this can be written down by analogy to \S1 of the
Supplemental Material for \Refcite{WA12b}, or Eq.~(8) in
\Refcite{WA14}.  It is
\begin{equation}
\begin{split}
&W^n(\xvec')\,\refstatep^n(\xvec') =
{\textstyle\sum_{\xvec}}\> 
W^{n-1}(\xvec)\,\refstatep^{n-1}(\xvec)\\
&\hspace{7em}{}\times
\refp(\xvec\to\xvec') \times\frac{\targp(\xvec\to\xvec')}
{\refp(\xvec\to\xvec')}\,.
\end{split}
\label{eq:tq1}
\end{equation}
This looks formidable but can be taken apart piece by piece.  The
first factor in the sum on the right hand side is the total weight in
microstate $\xvec$ after $n-1$ steps.  The second factor is the
probability $\refp(\xvec\to\xvec')$ that this weight subsequently
propagates to $\xvec'$. The third factor arises because the weight is
updated at each step by multiplying by the relative transition
probabilities $\targp/\refp$. Cancelling factors of $\refp$,
\Eqref{eq:tq1} simplifies to
\begin{equation}
\begin{split}
W^n(\xvec')\,\refstatep^n(\xvec') 
&={\textstyle\sum_{\xvec}}\> 
W^{n-1}(\xvec)\,\refstatep^{n-1}(\xvec)\\
&\qquad\qquad{}\times \targp(\xvec\to\xvec') \,.
\end{split}
\label{eq:tq1b}
\end{equation}
This can be compared to the evolution equation for
$\targstatep^n(\xvec)$, which is
\begin{equation}
\targstatep^n(\xvec')
={\textstyle\sum_{\xvec}}\>
\targstatep^{n-1}(\xvec)
\times{\targp(\xvec\to\xvec')}\,.
\label{eq:tq1c}
\end{equation}
This simply expresses the fact that $\targp(\xvec\to\xvec')$ plays the
role of a transition probability matrix.  On inspection,
\Eqsref{eq:tq1b} and \eqref{eq:tq1c} are structurally identical, and
(unless we have chosen a pathological case) converge to
\emph{identical} steady state solutions to within a multiplicative
constant, irrespective of the choice of initial conditions.  This then
implies
\begin{equation}
\lim_{n\to\infty}
W^n(\xvec)\,\refstatep^n(\xvec) 
\propto\lim_{n\to\infty}\targstatep^n(\xvec)\,,
\end{equation}
or, recalling the definition in \Eqref{eq:q1},
\begin{equation}
\lim_{n\to\infty}\avrefbig{\delta(\xvec-\xvec_n) \times 
\frac{\targtrajp^n}{\reftrajp^n}}
\propto\frac{\targstatep^\infty(\xvec)}%
            {\refstatep^\infty(\xvec)}\,.\label{eq:blah}
\end{equation}
This proves the intended result, \Eqref{eq:10neq} in the main text.

As a corollary to all this, we note that the convergence to steady
state in \Eqsref{eq:tq1b} and \eqref{eq:tq1c} is governed by the
spectral properties of $\targp(\xvec\to\xvec')$, that is to say by the
\emph{target} system, not the reference system.  The implications of
this are discussed in the main text.

\section{Reweighting algorithm for birth-death process}\label{app:alg}
We simulate the birth-death process in \Eqref{eq:bd} using the
Gilespie kinetic Monte-Carlo algorithm \cite{Gil77}.  At some point in
time, suppose there are $x$ copies of $\A$.  The transition rates (the
reaction `propensities') are
\begin{equation}
a_\pm=\Bigl\{\begin{array}{ll}
\lambda & \quad x\to x+1\,,\\
x\mu & \quad x\to x-1\,.
\end{array}
\end{equation}
Accordingly, we select a time step $\delta t$ from an exponential
distribution $p(\delta t)=\atot e^{-\atot\delta t}$ where
$\atot=a_++a_-$, and reaction channel $x\to x\pm1$ with probabilities
$a_\pm/a$.  We advance time by $\delta t$ and update $x$ according to
the chosen reaction channel.  This completes one update step.

The transition probability for this update step is
\begin{equation}
p_\pm=({a_\pm}/{a})\times a e^{-a\,\delta t}=a_\pm e^{-a\,\delta t}\,.
\end{equation}
We omit the `measure' $d(\delta t)$, which in any case cancels out
below.

Now suppose we are simulating the reference system with death
rate $\refmu$, and we wish to reweight to a target system with a
different death rate $\targmu$.  The ratio of transition
probabilities is then, specifically,
\begin{equation}
\frac{\targp}{\refp} = \Bigl\{\begin{array}{ll}
e^{x(\refmu-\targmu)\delta t} & x\to x+1\,,\\
(\targmu/\refmu)\,e^{x(\refmu-\targmu)\delta t} & x\to x-1\,.\end{array}
\label{eq:pp0}
\end{equation}
Since $\lambda$ is the same in the target and reference systems, the
$\lambda$-dependence cancels in this.

Thus in order to reweight the system, we simulate the system using the
standard Gillespie algorithm, but after making the choice of time step
and reaction channel, we compute the ratio $\targp/\refp$ according to
\Eqref{eq:pp0}.  We maintain a circular history array of length $n$,
and keep a pointer into this history array.  When the value of
$\targp/\refp$ is calculated, it is stored in the array at the
location corresponding to the current value of the pointer, which is
then incremented (and reset to the beginning if it goes past the end
of the array).  The effect of this is that the oldest value of
$\targp/\refp$ is replaced by the most recent value, whilst keeping
all the intermediate values.

For steady state problems we typically (re)set time to $t=0$ and
iterate the above algorithm until $t>\tsamp$, where $\tsamp$ is some
sampling interval.  At that point we make a note of the state of the
system (\ie\ the value of $x$ in this case) and the reweighting factor
$W$, which is simply the product of the $\targp/\refp$ values stored
in the history array at that point in time.  We keep track of these
values of $x$ and $W$ and use them to compute the desired averages,
$\avref{W}$ and $\avref{xW}$.

The value of $\tsamp$ should be chosen so that the system is fully
relaxed between samples.  This ensures statistical independence of the
sampling points.  Additionally, we should make sure $\tsamp$ is
sufficiently large so that the history array is completely refreshed
between samples (this also has the side-effect of `pump-priming' the
array from a cold start).  In the present study we set $\tsamp$ equal
to the larger of $5\refmu^{-1}$ or $2\avref{\delta t}n$.

\section{Extreme value statistics and trajectory weights}\label{app:evs}
Suppose that we simulate $N$ trajectories of length $n$.  This
generates $N$ samples of the reweighting factor $W$, where $z=\ln W$
is normally distributed with some mean $m$ and variance $v$.  Although
we can get good coverage of the underlying normal distribution with
moderate value of $N$, the problem arises because we need to sample
the tail of the normal distribution when constructing averages
weighted with $W=e^z$.

The $W$-weighted averages we are trying to calculate are governed by
an overall multiplier of the form
\begin{equation}
e^z\times e^{-(z-m)^2/2v}=e^{z-(z-m)^2/2v}\,.\label{eq:mult}
\end{equation}
This comprises the weight $W=e^z$ multiplied by the distribution of
$z=\ln W$ values (\ie\ the same as $W\times P(\ln W)$ in
\Figref{fig:plw}).  \Eqref{eq:mult} has a maximum at $z=m+v$.  To
achieve good statistics we need to sample the tail of the
$z$-distribution in this vicinity.  This is a very stringent
requirement, and in particular is not simply satisfied by reaching out
to some multiple of the standard deviation $\sqrt{v}$ beyond the mean.

This consideration can be expressed in another way by considering the
extreme value statistics of $z$ \cite{Gum58, Col01}.  Recall that for
$N$ samples of a normal distribution, the typical maximum value
reached (the `high water mark') is $\zmax\approx m+\sqrt{v\ln N}$.  If
we demand that $\zmax\agt m+v$, then at least the `high water mark' of
the sampled $z$ values is beyond the peak in \Eqref{eq:mult}.  This
translates into $\sqrt{v\ln N}\agt v$, or $N\agt e^v$.  Therefore
$N\agt e^{\beta n}$ for $n\gg1$, with some coefficient $\beta$.
Turning this around it means that the relevant trajectories which
contribute to the weighted averages become exponentially rare as $n$
increases, supporting the claim made in the main text.


\begin{thebibliography}{50}%
\makeatletter
\providecommand \@ifxundefined [1]{%
 \@ifx{#1\undefined}
}%
\providecommand \@ifnum [1]{%
 \ifnum #1\expandafter \@firstoftwo
 \else \expandafter \@secondoftwo
 \fi
}%
\providecommand \@ifx [1]{%
 \ifx #1\expandafter \@firstoftwo
 \else \expandafter \@secondoftwo
 \fi
}%
\providecommand \natexlab [1]{#1}%
\providecommand \enquote  [1]{``#1''}%
\providecommand \bibnamefont  [1]{#1}%
\providecommand \bibfnamefont [1]{#1}%
\providecommand \citenamefont [1]{#1}%
\providecommand \href@noop [0]{\@secondoftwo}%
\providecommand \href [0]{\begingroup \@sanitize@url \@href}%
\providecommand \@href[1]{\@@startlink{#1}\@@href}%
\providecommand \@@href[1]{\endgroup#1\@@endlink}%
\providecommand \@sanitize@url [0]{\catcode `\\12\catcode `\$12\catcode
  `\&12\catcode `\#12\catcode `\^12\catcode `\_12\catcode `\%12\relax}%
\providecommand \@@startlink[1]{}%
\providecommand \@@endlink[0]{}%
\providecommand \url  [0]{\begingroup\@sanitize@url \@url }%
\providecommand \@url [1]{\endgroup\@href {#1}{\urlprefix }}%
\providecommand \urlprefix  [0]{URL }%
\providecommand \Eprint [0]{\href }%
\providecommand \doibase [0]{http://dx.doi.org/}%
\providecommand \selectlanguage [0]{\@gobble}%
\providecommand \bibinfo  [0]{\@secondoftwo}%
\providecommand \bibfield  [0]{\@secondoftwo}%
\providecommand \translation [1]{[#1]}%
\providecommand \BibitemOpen [0]{}%
\providecommand \bibitemStop [0]{}%
\providecommand \bibitemNoStop [0]{.\EOS\space}%
\providecommand \EOS [0]{\spacefactor3000\relax}%
\providecommand \BibitemShut  [1]{\csname bibitem#1\endcsname}%
\let\auto@bib@innerbib\@empty
\bibitem [{\citenamefont {Frenkel}\ and\ \citenamefont {Smit}(2002)}]{FS02}%
  \BibitemOpen
  \bibfield  {author} {\bibinfo {author} {\bibfnamefont {D.}~\bibnamefont
  {Frenkel}}\ and\ \bibinfo {author} {\bibfnamefont {B.}~\bibnamefont {Smit}},\
  }\href@noop {} {\emph {\bibinfo {title} {Understanding {M}olecular
  {S}imulation: from {A}lgorithms to {A}pplications}}},\ \bibinfo {edition}
  {2nd}\ ed.\ (\bibinfo  {publisher} {Academic},\ \bibinfo {address} {San
  Diego, California},\ \bibinfo {year} {2002})\BibitemShut {NoStop}%
\bibitem [{\citenamefont {Torrie}\ and\ \citenamefont {Valleau}(1977)}]{TV77}%
  \BibitemOpen
  \bibfield  {author} {\bibinfo {author} {\bibfnamefont {G.}~\bibnamefont
  {Torrie}}\ and\ \bibinfo {author} {\bibfnamefont {J.}~\bibnamefont
  {Valleau}},\ }\href@noop {} {\bibfield  {journal} {\bibinfo  {journal} {J.
  Comp. Phys.}\ }\textbf {\bibinfo {volume} {23}},\ \bibinfo {pages} {187}
  (\bibinfo {year} {1977})}\BibitemShut {NoStop}%
\bibitem [{\citenamefont {Evans}\ \emph {et~al.}(1995)\citenamefont {Evans},
  \citenamefont {Foster}, \citenamefont {Godr\`eche},\ and\ \citenamefont
  {Mukamel}}]{EFG95}%
  \BibitemOpen
  \bibfield  {author} {\bibinfo {author} {\bibfnamefont {M.~R.}\ \bibnamefont
  {Evans}}, \bibinfo {author} {\bibfnamefont {D.~P.}\ \bibnamefont {Foster}},
  \bibinfo {author} {\bibfnamefont {C.}~\bibnamefont {Godr\`eche}}, \ and\
  \bibinfo {author} {\bibfnamefont {D.}~\bibnamefont {Mukamel}},\ }\href@noop
  {} {\bibfield  {journal} {\bibinfo  {journal} {Phys. Rev. Lett.}\ }\textbf
  {\bibinfo {volume} {74}},\ \bibinfo {pages} {208} (\bibinfo {year}
  {1995})}\BibitemShut {NoStop}%
\bibitem [{\citenamefont {Stratford}\ \emph {et~al.}(2005)\citenamefont
  {Stratford}, \citenamefont {Adhikari}, \citenamefont {Pagonabarraga},
  \citenamefont {Desplat},\ and\ \citenamefont {Cates}}]{Stratford2005}%
  \BibitemOpen
  \bibfield  {author} {\bibinfo {author} {\bibfnamefont {K.}~\bibnamefont
  {Stratford}}, \bibinfo {author} {\bibfnamefont {R.}~\bibnamefont {Adhikari}},
  \bibinfo {author} {\bibfnamefont {I.}~\bibnamefont {Pagonabarraga}}, \bibinfo
  {author} {\bibfnamefont {J.-C.}\ \bibnamefont {Desplat}}, \ and\ \bibinfo
  {author} {\bibfnamefont {M.~E.}\ \bibnamefont {Cates}},\ }\href@noop {}
  {\bibfield  {journal} {\bibinfo  {journal} {Science}\ }\textbf {\bibinfo
  {volume} {309}},\ \bibinfo {pages} {2198} (\bibinfo {year}
  {2005})}\BibitemShut {NoStop}%
\bibitem [{\citenamefont {Arkin}\ \emph {et~al.}(1998)\citenamefont {Arkin},
  \citenamefont {Ross},\ and\ \citenamefont {McAdams}}]{ARM98}%
  \BibitemOpen
  \bibfield  {author} {\bibinfo {author} {\bibfnamefont {A.}~\bibnamefont
  {Arkin}}, \bibinfo {author} {\bibfnamefont {J.}~\bibnamefont {Ross}}, \ and\
  \bibinfo {author} {\bibfnamefont {H.~H.}\ \bibnamefont {McAdams}},\
  }\href@noop {} {\bibfield  {journal} {\bibinfo  {journal} {Genetics}\
  }\textbf {\bibinfo {volume} {149}},\ \bibinfo {pages} {1633} (\bibinfo {year}
  {1998})}\BibitemShut {NoStop}%
\bibitem [{\citenamefont {Santill\'an}\ and\ \citenamefont
  {Mackey}(2004)}]{SM04}%
  \BibitemOpen
  \bibfield  {author} {\bibinfo {author} {\bibfnamefont {M.}~\bibnamefont
  {Santill\'an}}\ and\ \bibinfo {author} {\bibfnamefont {M.~C.}\ \bibnamefont
  {Mackey}},\ }\href@noop {} {\bibfield  {journal} {\bibinfo  {journal}
  {Biophys J.}\ }\textbf {\bibinfo {volume} {86}},\ \bibinfo {pages} {75}
  (\bibinfo {year} {2004})}\BibitemShut {NoStop}%
\bibitem [{\citenamefont {Morelli}\ \emph {et~al.}(2009)\citenamefont
  {Morelli}, \citenamefont {ten Wolde},\ and\ \citenamefont {Allen}}]{MtW09}%
  \BibitemOpen
  \bibfield  {author} {\bibinfo {author} {\bibfnamefont {M.~J.}\ \bibnamefont
  {Morelli}}, \bibinfo {author} {\bibfnamefont {P.~R.}\ \bibnamefont {ten
  Wolde}}, \ and\ \bibinfo {author} {\bibfnamefont {R.~J.}\ \bibnamefont
  {Allen}},\ }\href@noop {} {\bibfield  {journal} {\bibinfo  {journal} {Proc.
  Natl. Acad. Sci. USA}\ }\textbf {\bibinfo {volume} {106}},\ \bibinfo {pages}
  {8101} (\bibinfo {year} {2009})}\BibitemShut {NoStop}%
\bibitem [{\citenamefont {Warren}\ and\ \citenamefont {{ten
  Wolde}}(2005)}]{WtW05}%
  \BibitemOpen
  \bibfield  {author} {\bibinfo {author} {\bibfnamefont {P.~B.}\ \bibnamefont
  {Warren}}\ and\ \bibinfo {author} {\bibfnamefont {P.~R.}\ \bibnamefont {{ten
  Wolde}}},\ }\href@noop {} {\bibfield  {journal} {\bibinfo  {journal} {J.
  Phys. Chem. B}\ }\textbf {\bibinfo {volume} {109}},\ \bibinfo {pages} {6812}
  (\bibinfo {year} {2005})}\BibitemShut {NoStop}%
\bibitem [{\citenamefont {Warmflash}\ \emph {et~al.}(2007)\citenamefont
  {Warmflash}, \citenamefont {Bhimalapuram},\ and\ \citenamefont
  {Dinner}}]{WBD07}%
  \BibitemOpen
  \bibfield  {author} {\bibinfo {author} {\bibfnamefont {A.}~\bibnamefont
  {Warmflash}}, \bibinfo {author} {\bibfnamefont {P.}~\bibnamefont
  {Bhimalapuram}}, \ and\ \bibinfo {author} {\bibfnamefont {A.~R.}\
  \bibnamefont {Dinner}},\ }\href@noop {} {\bibfield  {journal} {\bibinfo
  {journal} {J. Chem. Phys.}\ }\textbf {\bibinfo {volume} {127}},\ \bibinfo
  {pages} {154112} (\bibinfo {year} {2007})}\BibitemShut {NoStop}%
\bibitem [{\citenamefont {Valeriani}\ \emph {et~al.}(2007)\citenamefont
  {Valeriani}, \citenamefont {Allen}, \citenamefont {Morelli}, \citenamefont
  {Frenkel},\ and\ \citenamefont {ten Wolde}}]{VAM+07}%
  \BibitemOpen
  \bibfield  {author} {\bibinfo {author} {\bibfnamefont {C.}~\bibnamefont
  {Valeriani}}, \bibinfo {author} {\bibfnamefont {R.~J.}\ \bibnamefont
  {Allen}}, \bibinfo {author} {\bibfnamefont {M.~J.}\ \bibnamefont {Morelli}},
  \bibinfo {author} {\bibfnamefont {D.}~\bibnamefont {Frenkel}}, \ and\
  \bibinfo {author} {\bibfnamefont {P.~R.}\ \bibnamefont {ten Wolde}},\
  }\href@noop {} {\bibfield  {journal} {\bibinfo  {journal} {J. Chem. Phys.}\
  }\textbf {\bibinfo {volume} {127}},\ \bibinfo {pages} {114109} (\bibinfo
  {year} {2007})}\BibitemShut {NoStop}%
\bibitem [{\citenamefont {Dickson}\ \emph
  {et~al.}(2009{\natexlab{a}})\citenamefont {Dickson}, \citenamefont
  {Warmflash},\ and\ \citenamefont {Dinner}}]{DWD09a}%
  \BibitemOpen
  \bibfield  {author} {\bibinfo {author} {\bibfnamefont {A.}~\bibnamefont
  {Dickson}}, \bibinfo {author} {\bibfnamefont {A.}~\bibnamefont {Warmflash}},
  \ and\ \bibinfo {author} {\bibfnamefont {A.~R.}\ \bibnamefont {Dinner}},\
  }\href@noop {} {\bibfield  {journal} {\bibinfo  {journal} {J. Chem. Phys.}\
  }\textbf {\bibinfo {volume} {130}},\ \bibinfo {pages} {074104} (\bibinfo
  {year} {2009}{\natexlab{a}})}\BibitemShut {NoStop}%
\bibitem [{\citenamefont {Dickson}\ \emph
  {et~al.}(2009{\natexlab{b}})\citenamefont {Dickson}, \citenamefont
  {Warmflash},\ and\ \citenamefont {Dinner}}]{DWD09b}%
  \BibitemOpen
  \bibfield  {author} {\bibinfo {author} {\bibfnamefont {A.}~\bibnamefont
  {Dickson}}, \bibinfo {author} {\bibfnamefont {A.}~\bibnamefont {Warmflash}},
  \ and\ \bibinfo {author} {\bibfnamefont {A.~R.}\ \bibnamefont {Dinner}},\
  }\href@noop {} {\bibfield  {journal} {\bibinfo  {journal} {J. Chem. Phys.}\
  }\textbf {\bibinfo {volume} {131}},\ \bibinfo {pages} {154104} (\bibinfo
  {year} {2009}{\natexlab{b}})}\BibitemShut {NoStop}%
\bibitem [{\citenamefont {Dickson}\ and\ \citenamefont {Dinner}(2010)}]{DD10}%
  \BibitemOpen
  \bibfield  {author} {\bibinfo {author} {\bibfnamefont {A.}~\bibnamefont
  {Dickson}}\ and\ \bibinfo {author} {\bibfnamefont {A.~R.}\ \bibnamefont
  {Dinner}},\ }\href@noop {} {\bibfield  {journal} {\bibinfo  {journal} {Annu.
  Rev. Phys. Chem.}\ }\textbf {\bibinfo {volume} {61}},\ \bibinfo {pages} {441}
  (\bibinfo {year} {2010})}\BibitemShut {NoStop}%
\bibitem [{\citenamefont {Allen}\ \emph {et~al.}(2009)\citenamefont {Allen},
  \citenamefont {Valeriani},\ and\ \citenamefont {ten Wolde}}]{Allen2009}%
  \BibitemOpen
  \bibfield  {author} {\bibinfo {author} {\bibfnamefont {R.~J.}\ \bibnamefont
  {Allen}}, \bibinfo {author} {\bibfnamefont {C.}~\bibnamefont {Valeriani}}, \
  and\ \bibinfo {author} {\bibfnamefont {P.~R.}\ \bibnamefont {ten Wolde}},\
  }\href@noop {} {\bibfield  {journal} {\bibinfo  {journal} {J. Phys. Condens.
  Matter}\ }\textbf {\bibinfo {volume} {21}},\ \bibinfo {pages} {463102}
  (\bibinfo {year} {2009})}\BibitemShut {NoStop}%
\bibitem [{\citenamefont {Becker}\ \emph {et~al.}(2012)\citenamefont {Becker},
  \citenamefont {Allen},\ and\ \citenamefont {ten Wolde}}]{BAtW12}%
  \BibitemOpen
  \bibfield  {author} {\bibinfo {author} {\bibfnamefont {N.~B.}\ \bibnamefont
  {Becker}}, \bibinfo {author} {\bibfnamefont {R.~J.}\ \bibnamefont {Allen}}, \
  and\ \bibinfo {author} {\bibfnamefont {P.~R.}\ \bibnamefont {ten Wolde}},\
  }\href@noop {} {\bibfield  {journal} {\bibinfo  {journal} {J. Chem. Phys.}\
  }\textbf {\bibinfo {volume} {136}},\ \bibinfo {pages} {174118} (\bibinfo
  {year} {2012})}\BibitemShut {NoStop}%
\bibitem [{\citenamefont {Asmussen}\ and\ \citenamefont {Glynn}(2007)}]{AG07}%
  \BibitemOpen
  \bibfield  {author} {\bibinfo {author} {\bibfnamefont {S.}~\bibnamefont
  {Asmussen}}\ and\ \bibinfo {author} {\bibfnamefont {P.~W.}\ \bibnamefont
  {Glynn}},\ }\href@noop {} {\emph {\bibinfo {title} {Stochastic {S}imulation:
  {A}lgorithms and {A}nalysis}}}\ (\bibinfo  {publisher} {Springer},\ \bibinfo
  {address} {New York},\ \bibinfo {year} {2007})\BibitemShut {NoStop}%
\bibitem [{\citenamefont {Touchette}(2012)}]{Touchette2012}%
  \BibitemOpen
  \bibfield  {author} {\bibinfo {author} {\bibfnamefont {H.}~\bibnamefont
  {Touchette}},\ }\href@noop {} {\enquote {\bibinfo {title} {A basic
  introduction to large deviations: Theory, applications, simulations},}\ }
  (\bibinfo {year} {2012}),\ \bibinfo {note} {arxiv: 1103.4146v3}\BibitemShut
  {NoStop}%
\bibitem [{\citenamefont {Glasserman}(1990)}]{Gla90}%
  \BibitemOpen
  \bibfield  {author} {\bibinfo {author} {\bibfnamefont {P.}~\bibnamefont
  {Glasserman}},\ }\href@noop {} {\emph {\bibinfo {title} {Gradient
  {E}stimation via {P}erturbation {A}nalysis}}}\ (\bibinfo  {publisher}
  {Springer},\ \bibinfo {address} {New {Y}ork},\ \bibinfo {year}
  {1990})\BibitemShut {NoStop}%
\bibitem [{\citenamefont {Wang}\ and\ \citenamefont {Rathinam}(2016)}]{WR16}%
  \BibitemOpen
  \bibfield  {author} {\bibinfo {author} {\bibfnamefont {T.}~\bibnamefont
  {Wang}}\ and\ \bibinfo {author} {\bibfnamefont {M.}~\bibnamefont
  {Rathinam}},\ }\href@noop {} {\bibfield  {journal} {\bibinfo  {journal}
  {Siam/ASA J. Uncertainty Quant.}\ }\textbf {\bibinfo {volume} {4}},\ \bibinfo
  {pages} {1288} (\bibinfo {year} {2016})}\BibitemShut {NoStop}%
\bibitem [{\citenamefont {Jarzynski}(1997)}]{Jar97}%
  \BibitemOpen
  \bibfield  {author} {\bibinfo {author} {\bibfnamefont {C.}~\bibnamefont
  {Jarzynski}},\ }\href@noop {} {\bibfield  {journal} {\bibinfo  {journal}
  {Phys. Rev. Lett.}\ }\textbf {\bibinfo {volume} {78}},\ \bibinfo {pages}
  {2690} (\bibinfo {year} {1997})}\BibitemShut {NoStop}%
\bibitem [{\citenamefont {Crooks}(1999)}]{Cro99}%
  \BibitemOpen
  \bibfield  {author} {\bibinfo {author} {\bibfnamefont {G.~E.}\ \bibnamefont
  {Crooks}},\ }\href@noop {} {\bibfield  {journal} {\bibinfo  {journal} {Phys.
  Rev. E}\ }\textbf {\bibinfo {volume} {60}},\ \bibinfo {pages} {2721}
  (\bibinfo {year} {1999})}\BibitemShut {NoStop}%
\bibitem [{\citenamefont {Sun}(2003)}]{Sun03}%
  \BibitemOpen
  \bibfield  {author} {\bibinfo {author} {\bibfnamefont {S.~X.}\ \bibnamefont
  {Sun}},\ }\href@noop {} {\bibfield  {journal} {\bibinfo  {journal} {J. Chem.
  Phys.}\ }\textbf {\bibinfo {volume} {118}},\ \bibinfo {pages} {5769}
  (\bibinfo {year} {2003})}\BibitemShut {NoStop}%
\bibitem [{\citenamefont {Warren}\ and\ \citenamefont
  {Allen}(2012{\natexlab{a}})}]{WA12}%
  \BibitemOpen
  \bibfield  {author} {\bibinfo {author} {\bibfnamefont {P.~B.}\ \bibnamefont
  {Warren}}\ and\ \bibinfo {author} {\bibfnamefont {R.~J.}\ \bibnamefont
  {Allen}},\ }\href@noop {} {\bibfield  {journal} {\bibinfo  {journal} {J.
  Chem. Phys.}\ }\textbf {\bibinfo {volume} {136}},\ \bibinfo {pages} {104106}
  (\bibinfo {year} {2012}{\natexlab{a}})}\BibitemShut {NoStop}%
\bibitem [{\citenamefont {Warren}\ and\ \citenamefont
  {Allen}(2012{\natexlab{b}})}]{WA12b}%
  \BibitemOpen
  \bibfield  {author} {\bibinfo {author} {\bibfnamefont {P.~B.}\ \bibnamefont
  {Warren}}\ and\ \bibinfo {author} {\bibfnamefont {R.~J.}\ \bibnamefont
  {Allen}},\ }\href@noop {} {\bibfield  {journal} {\bibinfo  {journal} {Phys.
  Rev. Lett.}\ }\textbf {\bibinfo {volume} {109}},\ \bibinfo {pages} {250601}
  (\bibinfo {year} {2012}{\natexlab{b}})}\BibitemShut {NoStop}%
\bibitem [{\citenamefont {Warren}\ and\ \citenamefont {Allen}(2014)}]{WA14}%
  \BibitemOpen
  \bibfield  {author} {\bibinfo {author} {\bibfnamefont {P.~B.}\ \bibnamefont
  {Warren}}\ and\ \bibinfo {author} {\bibfnamefont {R.~J.}\ \bibnamefont
  {Allen}},\ }\href@noop {} {\bibfield  {journal} {\bibinfo  {journal}
  {Entropy}\ }\textbf {\bibinfo {volume} {16}},\ \bibinfo {pages} {221}
  (\bibinfo {year} {2014})}\BibitemShut {NoStop}%
\bibitem [{\citenamefont {Onsager}\ and\ \citenamefont {Machlup}(1953)}]{OM53}%
  \BibitemOpen
  \bibfield  {author} {\bibinfo {author} {\bibfnamefont {L.}~\bibnamefont
  {Onsager}}\ and\ \bibinfo {author} {\bibfnamefont {S.}~\bibnamefont
  {Machlup}},\ }\href@noop {} {\bibfield  {journal} {\bibinfo  {journal} {Phys.
  Rev.}\ }\textbf {\bibinfo {volume} {91}},\ \bibinfo {pages} {1505} (\bibinfo
  {year} {1953})}\BibitemShut {NoStop}%
\bibitem [{\citenamefont {Adib}(2008)}]{Adi08}%
  \BibitemOpen
  \bibfield  {author} {\bibinfo {author} {\bibfnamefont {A.~B.}\ \bibnamefont
  {Adib}},\ }\href@noop {} {\bibfield  {journal} {\bibinfo  {journal} {J. Phys.
  Chem. B}\ }\textbf {\bibinfo {volume} {112}},\ \bibinfo {pages} {5910}
  (\bibinfo {year} {2008})}\BibitemShut {NoStop}%
\bibitem [{\citenamefont {Harland}\ and\ \citenamefont {Sun}(2007)}]{HS07}%
  \BibitemOpen
  \bibfield  {author} {\bibinfo {author} {\bibfnamefont {B.}~\bibnamefont
  {Harland}}\ and\ \bibinfo {author} {\bibfnamefont {S.~X.}\ \bibnamefont
  {Sun}},\ }\href@noop {} {\bibfield  {journal} {\bibinfo  {journal} {J. Chem.
  Phys.}\ }\textbf {\bibinfo {volume} {127}},\ \bibinfo {pages} {104103}
  (\bibinfo {year} {2007})}\BibitemShut {NoStop}%
\bibitem [{\citenamefont {Kuwahara}\ and\ \citenamefont {Mura}(2008)}]{KM08}%
  \BibitemOpen
  \bibfield  {author} {\bibinfo {author} {\bibfnamefont {H.}~\bibnamefont
  {Kuwahara}}\ and\ \bibinfo {author} {\bibfnamefont {I.}~\bibnamefont
  {Mura}},\ }\href@noop {} {\bibfield  {journal} {\bibinfo  {journal} {J. Chem.
  Phys.}\ }\textbf {\bibinfo {volume} {129}},\ \bibinfo {pages} {165101}
  (\bibinfo {year} {2008})}\BibitemShut {NoStop}%
\bibitem [{\citenamefont {Gillespie}\ \emph {et~al.}(2009)\citenamefont
  {Gillespie}, \citenamefont {Roh},\ and\ \citenamefont {Petzold}}]{GRP09}%
  \BibitemOpen
  \bibfield  {author} {\bibinfo {author} {\bibfnamefont {D.~T.}\ \bibnamefont
  {Gillespie}}, \bibinfo {author} {\bibfnamefont {M.}~\bibnamefont {Roh}}, \
  and\ \bibinfo {author} {\bibfnamefont {L.~R.}\ \bibnamefont {Petzold}},\
  }\href@noop {} {\bibfield  {journal} {\bibinfo  {journal} {J. Chem. Phys.}\
  }\textbf {\bibinfo {volume} {130}},\ \bibinfo {pages} {174103} (\bibinfo
  {year} {2009})}\BibitemShut {NoStop}%
\bibitem [{\citenamefont {Roh}\ \emph {et~al.}(2010)\citenamefont {Roh},
  \citenamefont {Gillespie},\ and\ \citenamefont {Petzold}}]{RGP10}%
  \BibitemOpen
  \bibfield  {author} {\bibinfo {author} {\bibfnamefont {M.~K.}\ \bibnamefont
  {Roh}}, \bibinfo {author} {\bibfnamefont {D.~T.}\ \bibnamefont {Gillespie}},
  \ and\ \bibinfo {author} {\bibfnamefont {L.~R.}\ \bibnamefont {Petzold}},\
  }\href@noop {} {\bibfield  {journal} {\bibinfo  {journal} {J. Chem. Phys.}\
  }\textbf {\bibinfo {volume} {133}},\ \bibinfo {pages} {174106} (\bibinfo
  {year} {2010})}\BibitemShut {NoStop}%
\bibitem [{\citenamefont {Daigle}\ \emph {et~al.}(2011)\citenamefont {Daigle},
  \citenamefont {Roh}, \citenamefont {Gillespie},\ and\ \citenamefont
  {Petzold}}]{DRG+11}%
  \BibitemOpen
  \bibfield  {author} {\bibinfo {author} {\bibfnamefont {B.~J.}\ \bibnamefont
  {Daigle}, \bibfnamefont {Jr.}}, \bibinfo {author} {\bibfnamefont {M.~K.}\
  \bibnamefont {Roh}}, \bibinfo {author} {\bibfnamefont {D.~T.}\ \bibnamefont
  {Gillespie}}, \ and\ \bibinfo {author} {\bibfnamefont {L.~R.}\ \bibnamefont
  {Petzold}},\ }\href@noop {} {\bibfield  {journal} {\bibinfo  {journal} {J.
  Chem. Phys.}\ }\textbf {\bibinfo {volume} {134}},\ \bibinfo {pages} {044110}
  (\bibinfo {year} {2011})}\BibitemShut {NoStop}%
\bibitem [{\citenamefont {Roh}\ \emph {et~al.}(2011)\citenamefont {Roh},
  \citenamefont {Daigle}, \citenamefont {Gillespie},\ and\ \citenamefont
  {Petzold}}]{RDG+11}%
  \BibitemOpen
  \bibfield  {author} {\bibinfo {author} {\bibfnamefont {M.~K.}\ \bibnamefont
  {Roh}}, \bibinfo {author} {\bibfnamefont {B.~J.}\ \bibnamefont {Daigle},
  \bibfnamefont {Jr.}}, \bibinfo {author} {\bibfnamefont {D.~T.}\ \bibnamefont
  {Gillespie}}, \ and\ \bibinfo {author} {\bibfnamefont {L.~R.}\ \bibnamefont
  {Petzold}},\ }\href@noop {} {\bibfield  {journal} {\bibinfo  {journal} {J.
  Chem. Phys.}\ }\textbf {\bibinfo {volume} {135}},\ \bibinfo {pages} {234108}
  (\bibinfo {year} {2011})}\BibitemShut {NoStop}%
\bibitem [{\citenamefont {Plyasunov}\ and\ \citenamefont {Arkin}(2007)}]{PA07}%
  \BibitemOpen
  \bibfield  {author} {\bibinfo {author} {\bibfnamefont {S.}~\bibnamefont
  {Plyasunov}}\ and\ \bibinfo {author} {\bibfnamefont {A.}~\bibnamefont
  {Arkin}},\ }\href@noop {} {\bibfield  {journal} {\bibinfo  {journal} {J.
  Comput. Phys.}\ }\textbf {\bibinfo {volume} {221}},\ \bibinfo {pages} {724}
  (\bibinfo {year} {2007})}\BibitemShut {NoStop}%
\bibitem [{\citenamefont {Rosenbluth}\ and\ \citenamefont
  {Rosenbluth}(1955)}]{RR55}%
  \BibitemOpen
  \bibfield  {author} {\bibinfo {author} {\bibfnamefont {M.~N.}\ \bibnamefont
  {Rosenbluth}}\ and\ \bibinfo {author} {\bibfnamefont {A.~W.}\ \bibnamefont
  {Rosenbluth}},\ }\href@noop {} {\bibfield  {journal} {\bibinfo  {journal} {J.
  Chem. Phys.}\ }\textbf {\bibinfo {volume} {23}},\ \bibinfo {pages} {356}
  (\bibinfo {year} {1955})}\BibitemShut {NoStop}%
\bibitem [{\citenamefont {Batoulis}\ and\ \citenamefont {Kremer}(1988)}]{BK88}%
  \BibitemOpen
  \bibfield  {author} {\bibinfo {author} {\bibfnamefont {J.}~\bibnamefont
  {Batoulis}}\ and\ \bibinfo {author} {\bibfnamefont {K.}~\bibnamefont
  {Kremer}},\ }\href@noop {} {\bibfield  {journal} {\bibinfo  {journal} {J.
  Phys. A: math. Gen.}\ }\textbf {\bibinfo {volume} {21}},\ \bibinfo {pages}
  {127} (\bibinfo {year} {1988})}\BibitemShut {NoStop}%
\bibitem [{\citenamefont {Grassberger}(1997)}]{Gra97}%
  \BibitemOpen
  \bibfield  {author} {\bibinfo {author} {\bibfnamefont {P.}~\bibnamefont
  {Grassberger}},\ }\href@noop {} {\bibfield  {journal} {\bibinfo  {journal}
  {Phys. Rev. E}\ }\textbf {\bibinfo {volume} {56}},\ \bibinfo {pages} {3682}
  (\bibinfo {year} {1997})}\BibitemShut {NoStop}%
\bibitem [{\citenamefont {Grassberger}(1999)}]{Gra99}%
  \BibitemOpen
  \bibfield  {author} {\bibinfo {author} {\bibfnamefont {P.}~\bibnamefont
  {Grassberger}},\ }\href@noop {} {\bibfield  {journal} {\bibinfo  {journal}
  {J. Chem. Phys.}\ }\textbf {\bibinfo {volume} {111}},\ \bibinfo {pages} {440}
  (\bibinfo {year} {1999})}\BibitemShut {NoStop}%
\bibitem [{\citenamefont {Feller}(1968)}]{Fel68}%
  \BibitemOpen
  \bibfield  {author} {\bibinfo {author} {\bibfnamefont {W.}~\bibnamefont
  {Feller}},\ }\href@noop {} {\emph {\bibinfo {title} {An introduction to
  {P}robability {T}heory and {I}ts {A}pplications}}},\ \bibinfo {edition}
  {3rd}\ ed.,\ Vol.~\bibinfo {volume} {1}\ (\bibinfo  {publisher} {Wiley},\
  \bibinfo {address} {New York},\ \bibinfo {year} {1968})\BibitemShut {NoStop}%
\bibitem [{\citenamefont {Aitchison}\ and\ \citenamefont {Brown}(1957)}]{AB57}%
  \BibitemOpen
  \bibfield  {author} {\bibinfo {author} {\bibfnamefont {J.}~\bibnamefont
  {Aitchison}}\ and\ \bibinfo {author} {\bibfnamefont {J.~A.~C.}\ \bibnamefont
  {Brown}},\ }\href@noop {} {\emph {\bibinfo {title} {The {L}ognormal
  {D}istribution}}}\ (\bibinfo  {publisher} {CUP},\ \bibinfo {address}
  {Cambridge},\ \bibinfo {year} {1957})\BibitemShut {NoStop}%
\bibitem [{\citenamefont {Morelli}\ \emph {et~al.}(2008)\citenamefont
  {Morelli}, \citenamefont {Tanase-{N}icola}, \citenamefont {Allen},\ and\
  \citenamefont {ten Wolde}}]{Morelli2008}%
  \BibitemOpen
  \bibfield  {author} {\bibinfo {author} {\bibfnamefont {M.~J.}\ \bibnamefont
  {Morelli}}, \bibinfo {author} {\bibfnamefont {S.}~\bibnamefont
  {Tanase-{N}icola}}, \bibinfo {author} {\bibfnamefont {R.~J.}\ \bibnamefont
  {Allen}}, \ and\ \bibinfo {author} {\bibfnamefont {P.~R.}\ \bibnamefont {ten
  Wolde}},\ }\href@noop {} {\bibfield  {journal} {\bibinfo  {journal} {Biophys.
  J.}\ }\textbf {\bibinfo {volume} {94}},\ \bibinfo {pages} {3413} (\bibinfo
  {year} {2008})}\BibitemShut {NoStop}%
\bibitem [{pri()}]{private}%
  \BibitemOpen
  \href@noop {} {}\bibinfo {note} {H. Touchette, private
  communication.}\BibitemShut {Stop}%
\bibitem [{\citenamefont {Gillespie}(1977)}]{Gil77}%
  \BibitemOpen
  \bibfield  {author} {\bibinfo {author} {\bibfnamefont {D.~T.}\ \bibnamefont
  {Gillespie}},\ }\href@noop {} {\bibfield  {journal} {\bibinfo  {journal} {J.
  Phys. Chem.}\ }\textbf {\bibinfo {volume} {81}},\ \bibinfo {pages} {2340}
  (\bibinfo {year} {1977})}\BibitemShut {NoStop}%
\bibitem [{del()}]{deltat}%
  \BibitemOpen
  \href@noop {} {}\bibinfo {note} {Note that $\avtarg{\delta
  t}=\sum_{x=0}^\infty \targstatep^\infty/(k+\targmu x)\approx0.526$ is only
  $3$\% different from $\avref{\delta t}\approx0.540$, whereas $\targmu$ and
  $\refmu$ differ by 40\%, so replacing $\avref{\delta t}$ by $\avtarg{\delta
  t}$ in \Eqref{eq:rate} would make an indiscernable difference to the goodness
  of the fit.}\BibitemShut {Stop}%
\bibitem [{wpl()}]{wplnw}%
  \BibitemOpen
  \href@noop {} {}\bibinfo {note} {Since we are dealing with probability
  distributions, $P(W)\,dW= P(\ln W)\,d(\ln W)$. Multiplying through by $W$
  shows that $W\times P(\ln W)$ is the correct quantity to think about in this
  context.}\BibitemShut {Stop}%
\bibitem [{poo()}]{poor}%
  \BibitemOpen
  \href@noop {} {}\bibinfo {note} {Note that poor sampling of $P(\ln W)$ is
  distinct from the problem that arises when there is poor overlap between the
  reference and biased probability distributions.}\BibitemShut {Stop}%
\bibitem [{\citenamefont {Prellberg}\ and\ \citenamefont
  {Krawczyk}(2004)}]{PK04}%
  \BibitemOpen
  \bibfield  {author} {\bibinfo {author} {\bibfnamefont {T.}~\bibnamefont
  {Prellberg}}\ and\ \bibinfo {author} {\bibfnamefont {J.}~\bibnamefont
  {Krawczyk}},\ }\href@noop {} {\bibfield  {journal} {\bibinfo  {journal}
  {Phys. Rev. Lett.}\ }\textbf {\bibinfo {volume} {92}},\ \bibinfo {pages}
  {120602} (\bibinfo {year} {2004})}\BibitemShut {NoStop}%
\bibitem [{\citenamefont {Minh}\ and\ \citenamefont {Chodera}(2009)}]{MC09}%
  \BibitemOpen
  \bibfield  {author} {\bibinfo {author} {\bibfnamefont {D.~D.~L.}\
  \bibnamefont {Minh}}\ and\ \bibinfo {author} {\bibfnamefont {J.~D.}\
  \bibnamefont {Chodera}},\ }\href@noop {} {\bibfield  {journal} {\bibinfo
  {journal} {J. Chem. Phys.}\ }\textbf {\bibinfo {volume} {131}},\ \bibinfo
  {pages} {134110} (\bibinfo {year} {2009})}\BibitemShut {NoStop}%
\bibitem [{\citenamefont {Gumbel}(1958)}]{Gum58}%
  \BibitemOpen
  \bibfield  {author} {\bibinfo {author} {\bibfnamefont {E.~J.}\ \bibnamefont
  {Gumbel}},\ }\href@noop {} {\emph {\bibinfo {title} {Statistics of
  {E}xtremes}}}\ (\bibinfo  {publisher} {Columbia},\ \bibinfo {address} {New
  York},\ \bibinfo {year} {1958})\BibitemShut {NoStop}%
\bibitem [{\citenamefont {Coles}(2001)}]{Col01}%
  \BibitemOpen
  \bibfield  {author} {\bibinfo {author} {\bibfnamefont {S.}~\bibnamefont
  {Coles}},\ }\href@noop {} {\emph {\bibinfo {title} {An {I}ntroduction to
  {S}tatistical {M}odeling of {E}xtreme {V}alues}}}\ (\bibinfo  {publisher}
  {Springer},\ \bibinfo {address} {London},\ \bibinfo {year}
  {2001})\BibitemShut {NoStop}%
\end{thebibliography}
\end{document}